\documentclass[
 aip,
 jcp,
amsfonts,
amsmath,
amssymb,
reprint,%
]{revtex4-1}

\usepackage{graphicx,epstopdf}
\usepackage{dcolumn}
\usepackage{bm}
\usepackage{amsfonts}
\usepackage{float}
\usepackage[T1]{fontenc}
\usepackage[utf8]{inputenc}
\usepackage{mathptmx}
\usepackage{etoolbox}
\usepackage{multirow}
\usepackage[colorlinks,linkcolor=blue,urlcolor=blue,citecolor=blue]{hyperref}

\makeatletter
\def\@email#1#2{%
 \endgroup
 \patchcmd{\titleblock@produce}
  {\frontmatter@RRAPformat}
  {\frontmatter@RRAPformat{\produce@RRAP{*#1\href{mailto:#2}{#2}}}\frontmatter@RRAPformat}
  {}{}
}%
\makeatother
\begin{document}


\title{Kohn-Sham accuracy from orbital-free density functional theory via $\Delta$-machine learning}

\author{Shashikant Kumar}
 \affiliation{College of Engineering, Georgia Institute of Technology, Atlanta, GA 30332, USA}
\author{Xin Jing}
\affiliation{College of Engineering, Georgia Institute of Technology, Atlanta, GA 30332, USA}
\affiliation{College of Computing, Georgia Institute of Technology, Atlanta, GA 30332, USA}
\author{John E. Pask}
\affiliation{Physics Division, Lawrence Livermore National Laboratory, Livermore, CA 94550, USA}
\author{Andrew J. Medford}
\affiliation{College of Engineering, Georgia Institute of Technology, Atlanta, GA 30332, USA}%
\author{Phanish Suryanarayana}
\email[Email: ]{phanish.suryanarayana@ce.gatech.edu}
\affiliation{College of Engineering, Georgia Institute of Technology, Atlanta, GA 30332, USA}
\affiliation{College of Computing, Georgia Institute of Technology, Atlanta, GA 30332, USA}

\date{\today}

\begin{abstract}
We present a $\Delta$-machine learning model for obtaining Kohn-Sham accuracy from orbital-free density functional theory (DFT) calculations. In particular, we employ a machine learned force field (MLFF) scheme based on the kernel method to capture the difference between Kohn-Sham and orbital-free DFT energies/forces. We implement this model in the context of on-the-fly molecular dynamics simulations, and study its accuracy,  performance, and sensitivity to parameters  for representative systems. We find that the formalism not only improves the accuracy of Thomas-Fermi-von Weizs{\"a}cker (TFW) orbital-free energies and forces by more than two orders of magnitude, but is also more accurate than MLFFs based solely on Kohn-Sham DFT, while being more efficient and less sensitive to model parameters. We apply the framework to study the structure of molten Al$_{0.88}$Si$_{0.12}$, the results suggesting no aggregation of Si atoms, in agreement with a previous Kohn-Sham study performed at an order of magnitude smaller length and time scales. 
\end{abstract}
\maketitle

\section{\label{sec:Introduction}Introduction}
Kohn-Sham density functional theory (DFT) \cite{kohn1965self, hohenberg1964inhomogeneous} is a widely used  ab initio method  that has a high accuracy to cost ratio relative to other such first principles methods. However, while less expensive than wavefunction-based methods, Kohn-Sham calculations are still associated with significant computational cost. In particular, they scale cubically with the number of atoms/electrons in the system, a consequence of the orthonormality constraint on the Kohn-Sham orbitals, which restricts the length and time scales accessible to such a rigorous quantum mechanical investigation. This bottleneck is particularly severe in molecular dynamics (MD) simulations, where the Kohn-Sham equations may need to be solved thousands or even hundreds of thousands of times to reach time scales relevant to the phenomena of interest \cite{burke2012perspective}. 

Orbital-free DFT \cite{ligneres2005introduction} represents a simplified version of Kohn-Sham DFT in which the orbital-dependent non-interacting kinetic energy functional is replaced by an explicit functional of the electron density. In so doing, the orbitals are removed from the formalism, thereby reducing the problem to the calculation of the  electron density alone, in the spirit of the Hohenberg-Kohn theorem \cite{hohenberg1964inhomogeneous}. This can also be interpreted as replacing the fictitious system of non-interacting fermions with a fictitious system of non-interacting bosons \cite{levy1984exact, jiang2021time}. The absence of orbitals in orbital-free DFT allows for linear scaling with system size, thereby circumventing the cubic scaling bottleneck inherent in Kohn-Sham DFT. However, an exact kinetic energy functional in terms of the density is currently unknown. Since the pioneering work of Thomas \cite{thomas1927calculation}, Fermi \cite{fermi1928statistische}, and von Weizs{\"a}cker \cite{weizsacker1935theorie}, resulting in the Thomas-Fermi-von Weizs{\"a}cker (TFW) functional, the lack of universality has motivated the development of a number of semilocal \cite{constantin2019performance, luo2018simple, constantin2018semilocal, francisco2021analysis, perdew2007laplacian, constantin2017modified}, nonlocal \cite{huang2010nonlocal, constantin2018nonlocal, ho2008analytic, karasiev2012issues, chacon1985nonlocal, mazin2022constructing, wang1992kinetic, wang1999orbital, carling2003orbital, zhou2005improving, ho2007energetics, huang2010nonlocal, mazin1998nonlocal, shao2021revised, mi2018nonlocal, xu2022nonlocal, shin2014enhanced}, and machine learned \cite{golub2019kinetic, alghadeer2021highly, fujinami2020orbital, del2023variational, remme2023kineticnet, ryczko2022toward, kumar2022accurate, meyer2020machine, snyder2013orbital}  kinetic energy functionals. However, while advances have been significant, kinetic energy functionals remain limited in accuracy and transferability, particularly for complex systems with significant inhomogeneity in bonding. Furthermore, the lack of orbitals prevents the use of state-of-the-art pseudopotentials that are available for Kohn-Sham DFT calculations, which further limits the accuracy of orbital-free DFT. Consequently, orbital-free DFT has been less widely used than Kohn-Sham DFT in practice.

$\Delta$-machine learning represents a powerful method for learning the difference in quantities that can be computed from different levels of theory. The effectiveness of such a strategy relies on the qualitative  features in the quantities of interest being captured by the lower level of theory, whereby the differences in the quantities between the two theories are smoother and typically more localized than the quantities (obtained from the higher level of theory)  themselves. $\Delta$-machine learning has already proven useful in a number of different schemes, including the development of machine learned force fields (MLFFs) \cite{schleder2019dft, poltavsky2021machine, unke2021machine} for the difference in energies and/or forces from the following lower level to higher level theories: Kohn-Sham DFT to coupled cluster \cite{bowman2022delta, bogojeski2020quantum, nandi2021delta, qu2021breaking}, Hartree-Fock to coupled cluster \cite{zaspel2018boosting, ramakrishnan2015big, qiao2020orbnet}, Kohn-Sham DFT to random phase approximation for the correlation energy \cite{liu2022phase, verdi2023quantum}, and force fields to coupled cluster \cite{qu2022delta, qu2022many, yu2023status}. However, the use of such a  technique for learning the difference between orbital-free and Kohn-Sham DFT has not been studied heretofore.

In this work, we present a  $\Delta$-machine learning scheme for obtaining Kohn-Sham  accuracy from orbital-free DFT calculations. In particular, we model the difference in  Kohn-Sham and orbital-free DFT energies/forces using a kernel method based MLFF scheme. We implement this model within the context of on-the-fly MD simulations, and study its accuracy,  performance, and sensitivity to parameters  for representative systems. We find that the formalism not only improves the accuracy of TFW orbital-free DFT energies and forces by more than two orders of magnitude, but is also more accurate than MLFFs developed directly for Kohn-Sham DFT, while being more efficient and less sensitive to the model parameters. We apply the framework to study the possibility of Si aggregation in molten Al$_{0.88}$Si$_{0.12}$.

The remainder of the manuscript is organized as follows. In Section~\ref{sec:Formulation}, we discuss the formulation for the $\Delta$-machine learning model between orbital-free and Kohn-Sham DFT. Next, we study the scheme's accuracy, performance, and sensitivity to model parameters in Section~\ref{sec:ResultsDiscussion}, where we also apply it  to study the structure of molten Al$_{0.88}$Si$_{0.12}$. Finally, we provide concluding remarks in Section~\ref{sec:Conclusions}. 

\section{\label{sec:Formulation}Formulation}

\subsection{$\Delta$-machine learning model}
The error in the energy computed by orbital-free DFT  can be defined as:
\begin{equation} 
    E_{\Delta} = E_{\rm KS} - E_{\rm OF} \,, 
    \label{Eq:EnergyDecomp}
\end{equation}
where $E_{\rm KS}$ and $E_{\rm OF}$ are the Kohn-Sham and orbital-free electronic ground state energies, respectively.  The  dependence of these energies on the atomic positions has been suppressed for notational simplicity, a strategy adopted henceforth for all other such quantities. Since the ground state electron density in orbital-free DFT is \emph{close} to Kohn-Sham DFT \cite{thapa2023assessing}, and the electrostatics have the same formalism in both orbital-free and Kohn-Sham DFT,  $E_{\Delta}$ is largely free from long-range effects, and can therefore be assumed to depend on only the local environment.  This makes $E_{\Delta}$ particularly well suited for machine learning, as  exploited in the MLFF model discussed below. 

Given any atomic configuration/structure, we assume that the error in the orbital-free DFT energy can be decomposed as \cite{bartok2010gaussian}: 
\begin{equation} 
    E_{\Delta} = \sum_{e=1}^{N_{e}} \sum_{i=1}^{N_A^e} \epsilon_i^e  = \sum_{e=1}^{N_{e}} \sum_{i=1}^{N_A^e} \sum_{t=1}^{N_{T}^{e}} \tilde{w}_{t}^{e} k\left(\mathbf{x}_{i}^{e}, \tilde{\mathbf{x}}_t^{e} \right)
    \label{Eq:EnergyDecompML}
\end{equation}
where $\epsilon_i^e$ are the atomic energies, $e$ is the index that runs over the distinct chemical elements, $i$ is the index that runs over the different atoms for each element type, $N_{e}$ is the total  number of elements, and $N_A^e$ is the total number of atoms of the  element indexed by $e$.  In addition,  $\tilde{w}_{t}^{e}$ are the weights, and $k(\mathbf{x}_{i}^{e}, \tilde{\mathbf{x}}_t^{e})$ is a kernel that measures the distance between the descriptor vectors $\mathbf{x}_{i}^{e}$ and $\tilde{\mathbf{x}}_t^{e}$, the latter belonging to the training dataset for the chemical element indexed by $e$, which has $N_T^e$ such descriptors.  We choose the polynomial kernel \cite{bartok2013representing}: 
\begin{equation} 
   k(\mathbf{x}_i^e, \tilde{\mathbf{x}}_t^e) =   \left( \frac{\mathbf{x}_i^e \cdot \tilde{\mathbf{x}}_t^e}{|\mathbf{x}_i^e| |\tilde{\mathbf{x}}_t^e|}  \right)^{\xi}\,,
   \label{Eq:SOAPKernel}
\end{equation}
where the exponent $\xi$ is a parameter used to control the sharpness of the kernel. In addition, we employ the Smooth Overlap of Atomic Positions (SOAP) formalism \cite{bartok2013representing}, where the rotationally invariant descriptors correspond to the power spectrum of the coefficients $c_i^e$:
\begin{equation} 
   [\mathbf{x}_i^e]_{n_1 n_2 l}^{e_1 e_{2}} = \sqrt{\frac{8 \pi^2}{2l+1}}\sum_{m=-l}^{l} c_{i, e_1, n_1 lm}^{e} c_{i, e_{2},  n_2 lm}^{e *}  \,,
   \label{Eq:SOAPDescr}
\end{equation}
where $e_1, e_{2} \in \{1,2, \ldots ,N_e\}$, $n_1, n_2 \in \{1, 2, \ldots, N_r\}$, and $l \in \{0,1,...,L_{\rm max}\}$ are used to index the components of the descriptor, a vector of length $N_r^2 N_e^2 (L_{\rm max}+1)$. The coefficients $c_i^e$ themselves arise during the basis expansion of the atom density for the atom positioned at  $\mathbf{r}_i^e := (r_{i_1}^e, r_{i_2}^e, r_{i_3}^e)$: 
\begin{equation} 
   b_{i,\tilde{e}}^e\left(\mathbf{r}- \mathbf{r}_i^e\right) = \sum_{l=0}^{L_{\rm max}} \sum_{m=-l}^{l} \sum_{n=1}^{N_r} c_{i, \tilde{e}, nlm}^e j_l(q_{nl} | \mathbf{r} - \mathbf{r}_i^e |) Y_{lm}(\mathbf{r} - \mathbf{r}_i^e)   \,, 
   \label{Eq:RhoDecomp}
\end{equation}
where $\tilde{e} \in \{1,...,N_e\}$, $N_r$ is the number of radial basis functions for each $l$, $L_{\rm max}$ is the maximum value of $l$,  $Y_{lm}$ are the spherical harmonics, and $j_l$ are the normalized spherical Bessel functions, with $q_{nl}$ chosen such that $j_l(q_{nl}R_{\rm cut}) =0$, $R_{\rm cut}$ being the region of influence for any atom, i.e.,  the distance within which the atoms are assumed to contribute to the atom density. Since the chosen basis is complete, it allows for systematic convergence. The atom density itself is calculated as \cite{bartok2010gaussian}:
\begin{equation} 
     b_{i,\tilde{e}}^e\left(\mathbf{r} - \mathbf{r}_i^e \right) = \sum_{j=1}^{N_A^{\Tilde{e}}} f_{\rm cut}\left(|\mathbf{r}_{j}^{\tilde{e}} - \mathbf{r}_i^e|\right) g\left(\mathbf{r}- \mathbf{r}_{j}^{\tilde{e}} +\mathbf{r}_i^e\right) \,,
 \label{Eq:AtomDensity}
\end{equation}
where $f_{\rm cut}$ is the cosine function \cite{behler2007generalized}, and $g$ is the Gaussian function:
\begin{equation} 
   g\left(\mathbf{r}- \mathbf{r}_{j}^{\tilde{e}} +\mathbf{r}_i^e\right) = \frac{1}{\sqrt{2\sigma_{a}^2\pi}} \exp \left( - \frac{|\mathbf{r}- \mathbf{r}_{j}^{\tilde{e}} +\mathbf{r}_i^e|^2}{2\sigma_{a}^2} \right)   \,, 
   \label{Eq:GaussianFc}
\end{equation}
with $\sigma_a$ being the standard deviation of the Gaussian. A smaller value of $\sigma_{a}$ is suitable for systems with significant heterogeneity, while a larger value of $\sigma_{a}$ is suitable for  systems that are homogeneous. On using the single center expansion of the  Gaussian function \cite{kaufmann1989single}, the coefficients $c_i^e$ can be written as \cite{bartok2013representing, jinnouchi2019fly}: 
\begin{equation} \label{Eq:cnlmCoeff}
c_{i, \tilde{e}, nlm}^e = \sum_{j=1}^{N_A^{\Tilde{e}}} h_{nl} \left( |\mathbf{r}_{j}^{\tilde{e}} - \mathbf{r}_i^e | \right) Y_{lm}^* \left( \mathbf{r}_{j}^{\tilde{e}} - 
 \mathbf{r}_i^e\right) \,,
\end{equation}
where the function $h_{nl}$ itself takes the form:
\begin{equation} \label{Eq:hnl}
h_{nl}(r) = C(r) \int_{0}^{R_{\rm cut}} j_l(q_{nl}r') \exp \left(- \frac{r'^2 + r^2}{2\sigma_{a}^2}\right) l_l\left(\frac{rr'}{\sigma_{a}^2} \right)r'^2 {\rm d}r'  \,,
\end{equation}
with $C(r) = 4\pi  f_{\rm cut}(r) / \sqrt{2\sigma_a^2\pi }$, and $l_l$ being the modified spherical Bessel function of the first kind.

The above described model for $E_{\Delta}$ also provides access to  derivatives with respect to atomic positions. In particular, the first-order derivative of  $E_{\Delta}$ with respect to atomic position provides a correction to the force computed by  orbital-free DFT:
\begin{align}
f_{\Delta, j_{\alpha}}^{\Tilde{e}} & = -\frac{\partial E_{\Delta}}{\partial r_{j_{\alpha}}^{\Tilde{e}} } \nonumber \\
& = -\sum_{e=1}^{N_{e}} \sum_{i=1}^{N_A^e} \sum_{t=1}^{N_{T}^{e}} \tilde{w}_{t}^{e}  \frac{\partial k\left( \mathbf{x}_{i}^{e}, \tilde{\mathbf{x}}_t^{e} \right)}{\partial  \mathbf{x}_{i}^{e} } \cdot  \frac{\partial \mathbf{x}_{i}^{e}}{\partial   r_{j_{\alpha}}^{\Tilde{e}} } \nonumber \\ 
& = \sum_{e=1}^{N_{e}} \sum_{i=1}^{N_A^e} \sum_{t=1}^{N_{T}^{e}} \tilde{w}_{t}^{e} \xi k\left( \mathbf{x}_{i}^{e}, \tilde{\mathbf{x}}_t^{e} \right) \mathbf{z}_{i,t}^e \cdot  \frac{\partial \mathbf{x}_{i}^{e}}{\partial   r_{j_{\alpha}}^{\Tilde{e}} } \,,
\label{Eq:ForceFinalML}
\end{align} 
where 
\begin{subequations}
\begin{align} 
& \mathbf{z}_{i,t}^e  = \left(\frac{\mathbf{x}_{i}^{e}}{\mathbf{x}_{i}^{e} \cdot \mathbf{x}_{i}^{e}} - \displaystyle \frac{\tilde{\mathbf{x}}_t^{e}}{\mathbf{x}_{i}^{e} \cdot \tilde{\mathbf{x}}_t^{e}} \right) \,, \label{Eq:z} \\ 
& \frac{\partial[\mathbf{x}_i^e]_{n_1 n_2 l}^{e_1 e_{2}}}{\partial   r_{j_{\alpha}}^{\Tilde{e}} } = c_l \sum_{m=-l}^{l} \left( \frac{\partial c_{i, e_1,  n_1 lm}^{e}}{\partial   r_{j_{\alpha}}^{\Tilde{e}}} c_{i,  e_{2}, n_2 lm}^{e *}  + c_{i, e_1, n_1 lm}^{e}\frac{\partial  c_{i, e_{2}, n_2 lm}^{e *}}{\partial   r_{j_{\alpha}}^{\Tilde{e}}}  \right)  \,.
\label{Eq:DescriptorDeriv}
\end{align}
\end{subequations}
Above, $c_l = \sqrt{8 \pi^2/(2l+1)}$, and the derivatives of $ c_{i}^{e}$ can be obtained to be: 
\begin{equation}
    \frac{\partial c_{i, \Tilde{e}, nlm}^{e}}{\partial   r_{j_{\alpha}}^{\Tilde{e}} }= 
\begin{cases}
    \frac{ (r_{j_{\alpha}}^{\Tilde{e}} - r_{i_{\alpha}}^e)}{|\mathbf{r}_{j}^{\tilde{e}} - \mathbf{r}_i^e|} h_{nl}' \left( |\mathbf{r}_{j}^{\tilde{e}} - \mathbf{r}_i^e|\right) Y_{lm}^* \left( \mathbf{r}_{j}^{\tilde{e}} - \mathbf{r}_i^e \right)  \\
 + h_{nl} \left( |\mathbf{r}_{j}^{\tilde{e}} - \mathbf{r}_i^e| \right)\frac{\partial Y_{lm}^* \left( \mathbf{r}_{j}^{\tilde{e}} - \mathbf{r}_i^e \right)} {\partial (r_{j_{\alpha}}^{\Tilde{e}} - r_{i_{\alpha}}^e)},& \text{if } j \neq i,\\
    -\sum_{k=1, k \neq i}^{N_A^{\Tilde{e}}} \frac{\partial c_{i, \Tilde{e}, nlm}^{e}}{\partial   r_{k_{\alpha}}^{\Tilde{e}} },              & \text{if } j = i,
\end{cases}
\label{Eq:CnlmDeriv}
\end{equation}
where $ h_{nl}'$ denotes the derivative of $h_{nl}$. Similarly, the correction to the stress tensor, which represents the first-order derivative of $E_{\Delta}$ with respect to the strain tensor/deformation gradient, has been presented in Appendix~\ref{App:StressTensor}.


\subsection{Training: Bayesian linear regression}
The expressions for the energy and atomic forces appearing in Eqs.~\ref{Eq:EnergyDecompML} and \ref{Eq:ForceFinalML}, respectively, can be written in matrix form as:
\begin{equation} 
 E_{\Delta}^s  = \sum_{e=1}^{N_e}{\bm K}_E^{s,e} {\bm w}^e \,, 
\,\,\,  {\bm f}_{\Delta}^{s}  = \sum_{e=1}^{N_e} {\bm K}_f^{s,e} {\bm w}^e \,, 
 \quad  s \in \{1,...,N_{st}\} \,,
\label{Eq:EqnsMatForm}
\end{equation}
where $s$ is an index introduced for denoting  the configuration/structure,  ${\bm w}^e \in \mathbb{R}^{N_T^e \times 1}$ is a vector of the weights,  ${\bm K}_E^{s,e} \in \mathbb{R}^{1\times N_T^e}$ is the covariance matrix for the energy, and ${\bm K}_f^{s,e} \in \mathbb{R}^{\left(3\sum_e N_A^e\right)\times N_T^e}$ is the covariance matrix for the forces. For numerical purposes, the energy and forces are dimensionless: 
\begin{equation}
 E^s_{\Delta} := \frac{E_{\Delta} - \mu_E}{\sigma_E} \,, \quad  {\bm f}^s_{\Delta} := \frac{{\bm f}_{\Delta}}{\sigma_f} \,, 
 \label{Eq:EnergyForcesNorm}
\end{equation}
where $\mu_E$ and  ${\sigma_E}$ are the mean and standard deviation of $E^s_{\Delta}$, respectively, and ${\sigma_f}$ is the standard deviation of ${\bm f}^s_{\Delta}$, all defined over the different structures. 

The model weights ${\bm w}^e$  need to be determined from the data available for training. To do so, we consider Bayesian linear regression, which involves minimizing the loss function \cite{bishop2006pattern}: 
\begin{align} 
\mathcal{L}\left( {\bm w} \right) &=  \frac{\beta}{2} \left(w_E^2 \left\|   {\bm y}_E - \sum_{e=1}^{N_e} {\bm K}_E^e {\bm w}^e   \right\|_2^2  \right. \nonumber \\
&+ \left. w_f^2 \left\|  {\bm y}_f - \frac{\sigma_E}{\sigma_f} \sum_{e=1}^{N_e} {\bm K}_f^e{\bm w}^e  \right\|^2_2 \right) -
\frac{\alpha}{2} \|{\bm w} \|_2^2 \,,
\label{Eq:LossFuncMatForm}
\end{align}
where
\begin{subequations} 
\begin{align} 
{\bm w} &= \left[{{\bm w}^1}; \, {{\bm w}^2}; \,  \ldots ; \,  \ {{\bm w}^{N_e}} \right] \\
{\bm K}_E^e &= [{{\bm K}_E^{1,e}}; \, \ {{\bm K}_E^{2,e}}; \,  \ldots ; \,  {{\bm K}_E^{N_{st},e}}] \,, \\
{\bm K}_f^e &= [{{\bm K}_f^{1,e}}; \, \ {{\bm K}_f^{2,e}}; \, \ldots ; \, {{\bm K}_f^{N_{st},e}}] \,, \\
 {\bm y}_E & = [E^1_{\Delta}; \, E^2_{\Delta}; \, \ldots ;  \, E^{N_{st}}_{\Delta}] \,, \\
{\bm y}_f & = [{{\bm f}^1_{\Delta}}; \, {{\bm f}^2_{\Delta}}; \, \ldots; \,  {{\bm f}^{N_{st}}_{\Delta}}] \,,
\end{align}
\end{subequations}
with the semicolon used as a delimiter between the different rows of the vector/matrix. Above, $\alpha$ and $\beta$ are parameters, while $w_E$ and $w_f$ are the weighting factors for the errors in the energy and forces, respectively. The optimized weights can then be written as:
\begin{align}
{\bm w} =\beta {\bm C}_{{\bm w}} \left( {\bm K}_E^{\rm T} {\bm y}_E + {\bm K}_f^{\rm T} {\bm y}_f \right) \,,
\end{align} 
where 
\begin{align}
{\bm C}_{{\bm w}} = \left( \alpha  \mathit{I} + \beta\left({\bm K}_E^{\rm T} {\bm K}_E + {\bm K}_f^{\rm T} {\bm K}_f\right) \right)^{-1} \,,
\end{align}
with the matrices  ${\bm K}_E \in \mathbb{R}^{N_{st} \times (\sum_e^{N_e} N_T^e)}$ and ${\bm K}_f \in \mathbb{R}^{N_{st} (3\sum_e^{N_e} N_A^e) \times (\sum_e^{N_e} N_T^e)}$ of the form:
\begin{subequations}
\begin{align} \label{Eq:Global_Ke_y}
 {\bm K}_E &= \left[w_E {{\bm K}_E^1}; \, w_E {{\bm K}_E^2}; \, \ldots ; \,  w_E {{\bm K}_E^{N_e}}\right]  \,, \\
 {\bm K}_f &= \left[w_f\frac{\sigma_E}{\sigma_f} {{\bm K}_f^1}; \, w_f\frac{\sigma_E}{\sigma_f} {{\bm K}_f^2}; \, \ldots; \, w_f\frac{\sigma_E}{\sigma_f} {{\bm K}_f^{N_e}} \right] \,.
\end{align}
\end{subequations}
These optimized weights can then be used in Eq.~\ref{Eq:EqnsMatForm} to predict the energy and forces for any new structure, after having calculated the structure's covariance matrices for the energy and forces. The uncertainty in the energy (${\bm \sigma}_{{\bm y}_E}$) and forces (${\bm \sigma}_{{\bm y}_f}$) so calculated can be estimated as:
\begin{subequations}
\begin{align}
 {\bm \sigma}^2_{{\bm y}_E} & = {\rm diag} \left( \frac{1}{\beta} +  {\bm K}_E {\bm C}_{{\bm w}} {\bm K}_E^{\rm T} \right) \,,  \\
   {\bm \sigma}^2_{{\bm y}_f} & = {\rm diag} \left(\frac{1}{\beta} +  {\bm K}_f {\bm C}_{{\bm w}} {\bm K}_f^{\rm T} \right)  \,,
\end{align}
\end{subequations}
where diag(.) refers to the diagonal of the matrix. 


\section{\label{sec:ResultsDiscussion}Results and Discussion}

\begin{figure}[!htbp]
        \includegraphics[width=0.9\linewidth]{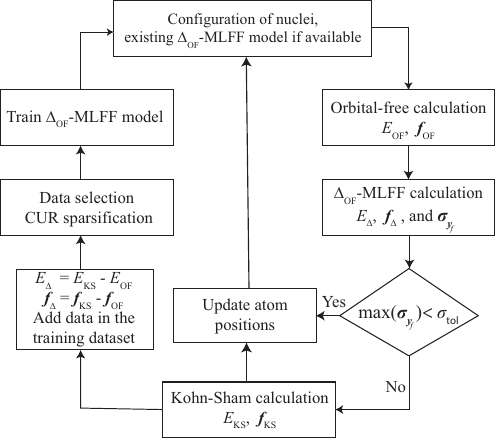}
        \caption{Outline of on-the-fly MD simulations using $\Delta_{\rm OF}$-MLFF.}
        \label{fig:flowchart}
\end{figure}

We have implemented the above described $\Delta$-machine learning formalism, henceforth referred to as $\Delta_{\rm OF}$-MLFF, within the SPARC electronic structure code  \cite{ghosh2017sparc1, ghosh2017sparc2, xu2021sparc, zhang2023version}. In particular, we have developed a framework for on-the-fly MD simulations, as outlined in Fig.~\ref{fig:flowchart}. In such simulations, after a few initial training MD steps for which Kohn-Sham calculations are performed, the energy and atomic forces are set to those predicted by the $\Delta_{\rm OF}$-MLFF model,  except when the uncertainty in the forces so computed  is larger than a specified threshold $\sigma_{\rm tol}$, at which point the energy and forces are set to those from a Kohn-Sham calculation, which are then also  included into the training dataset. Though the current work is focused on energy and atomic forces, the implementation has the capability to also include the stress tensor as part of the training and prediction. The threshold $\sigma_{\rm tol}$ is made to be adaptive by setting it to the maximum value of  ${\bm \sigma}_{{\bm y}_f}$ obtained in the $\Delta_{\rm OF}$-MLFF step that is subsequent to a training step \cite{jinnouchi2019fly}. Though this strategy results in increased number of Kohn-Sham calculations in the beginning of the MD simulation, the frequency of such calculations decreases rapidly, whereby the total number of Kohn-Sham calculations in the total MD simulation is generally lower than for other less stringent $\sigma_{\rm tol}$ choices. To avoid the cubic scaling bottleneck encountered in training, a two-step data selection procedure is employed: for any given configuration, we only add those atoms to the training dataset for which the Bayesian error in forces exceeds $\sigma_{\rm tol}$, and then perform   CUR  \cite{jinnouchi2019fly} on the resulting dataset for downsampling \cite{jinnouchi2019fly, briganti2023efficient, young2021transferable}. Note that the parameters $\alpha$ and $\beta$ are also dynamically updated within each training step,  by maximizing the evidence function of Bayesian linear regression \cite{bishop2006pattern}.

To study the sensitivity to parameters, accuracy, and efficiency of the $\Delta_{\rm OF}$-MLFF scheme, we consider the following systems: aluminum (Al) at 300 K, silicon (Si) at 300 K, and molten alloy of aluminum-silicon (Al$_{0.88}$Si$_{0.12}$) at 1473 K. We consider 108-, 216-, and 500-atom cells of Al, Si, and Al$_{0.88}$Si$_{0.12}$, respectively.  As an application, we also study Si aggregation in Al$_{0.88}$Si$_{0.12}$. In all instances, we perform isokinetic ensemble (NVK) MD with Gaussian thermostat \cite{minary2003algorithms}  and time steps of 4, 4, and 2 fs for the Al, Si, and  Al$_{0.88}$Si$_{0.12}$ systems, respectively. In the Kohn-Sham calculations, we employ  optimized norm-conserving Vanderbilt (ONCV) pseudopotentials \cite{hamann2013optimized} from the SPMS set \cite{shojaei2023soft}, and the following exchange-correlation functionals: local density approximation (LDA) \cite{perdew1981self}  for Al, and the Perdew-Burke-Ernzerhof (PBE) \cite{perdew1996generalized} variant of the generalized gradient approximation (GGA) for Si and Al$_{0.88}$Si$_{0.12}$. In the orbital-free calculations, which are also performed using a real-space formalism \cite{ghosh2016higher, suryanarayana2014augmented}, we employ the Thomas-Fermi-von Weizs{\"a}cker (TFW) kinetic energy functional \cite{werzsticxer1935theorie} (weight factor $\lambda=1$), LDA exchange-correlation functional, and   bulk-derived local pseudopotentials (BLPS)  \cite{zhou2004transferable}. Note that we have chosen a rather simple kinetic energy functional, since it suffices for achieving the target accuracy for the MD simulations in this work . Indeed, choosing more sophisticated kinetic energy functionals \cite{constantin2018nonlocal, luo2018simple, constantin2018semilocal, francisco2021analysis, perdew2007laplacian, constantin2017modified,  tomishima1966solution,  chacon1985nonlocal, mazin2022constructing, wang1992kinetic, wang1999orbital, carling2003orbital, zhou2005improving, ho2007energetics, huang2010nonlocal,  mazin1998nonlocal, shao2021revised, mi2018nonlocal, xu2022nonlocal, shin2014enhanced} is likely to further increase the accuracy of the $\Delta_{\rm OF}$-MLFF scheme, but will also substantially increase the cost. 

In presentation of the results, we compare against orbital-free DFT (OF-DFT) as well as MLFF learnt directly from Kohn-Sham DFT (KS-MLFF), i.e., same scheme used for machine learning $E_{\rm KS}$ and the corresponding atomic forces instead. Indeed, OF-DFT and KS-MLFF can be considered to be the limiting cases of the  $\Delta_{\rm OF}$-MLFF scheme. All errors are defined with respect to results obtained from Kohn-Sham DFT (KS-DFT). The energy error is defined to be the magnitude of the difference, and the force error is defined to be the maximum difference (in magnitude) in any force component among all the atoms. The errors are averaged over a sufficient number of MD steps to put statistical errors well below the errors under consideration.

\subsection{\label{Subsec:Sensitivity} Sensitivity to parameters}
We now study the sensitivity of $\Delta_{\rm OF}$-MLFF to the key parameters inherent to the model: hyperparameters in the machine learning model, and the discretization parameters in the orbital-free and Kohn-Sham calculations that are performed.   The focus here is on the atomic force errors, since the qualitative trends for the energy errors are similar, with the values generally being one order of magnitude smaller.

\begin{figure*}[!htbp]
        \includegraphics[width=0.8\linewidth]{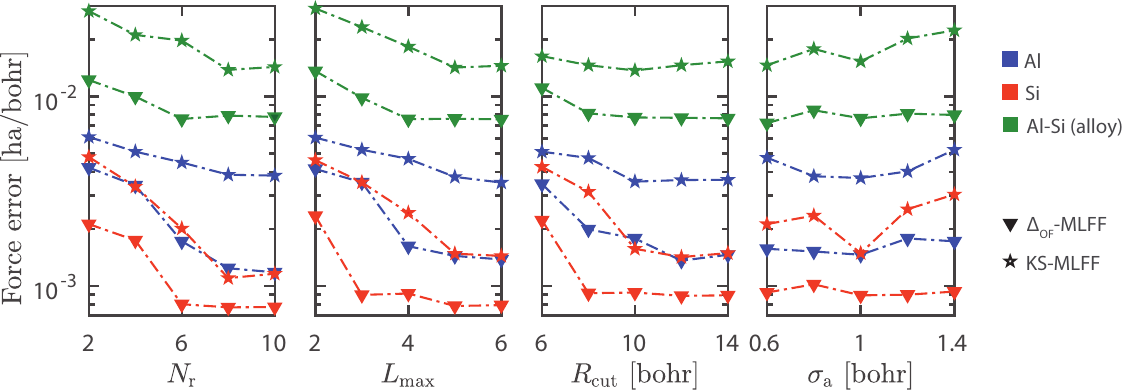}
        \caption{Variation in the force error  with respect to the model hyperparameters for the Al, Si, and Al$_{0.88}$Si$_{0.12}$ systems, averaged over 2000 MD steps.}
        \label{fig:scaling_error_MLFF}
\end{figure*}

First, we  study the variation in the $\Delta_{\rm OF}$-MLFF  errors   with respect to the hyperparameters $N_r$, $L_{\rm max}$, $R_{\rm max}$, and $\sigma_a$, the results so obtained are presented and compared with KS-MLFF in Fig.~\ref{fig:scaling_error_MLFF}.  The Kohn-Sham calculations employ $\Gamma$-point for Brillouin zone integration. Both orbital-free and Kohn-Sham calculations employ mesh sizes of $0.35$, $0.35$, and $0.32$ bohr for the Al, Si, and Al$_{0.88}$Si$_{0.12}$ systems, respectively, which translates to the energy and forces being converged to within $10^{-4}$ ha/bohr and $10^{-5}$ ha/atom, respectively. We observe that $\Delta_{\rm OF}$-MLFF has consistently smaller errors relative to KS-MLFF. Furthermore, the errors in $\Delta_{\rm OF}$-MLFF have significantly smaller variation with respect to $\sigma_a$, and saturate more rapidly with respect to the other hyperparameters,  leading to more efficient MLFF training/prediction as well as improved robustness and efficiency with respect to hyperparameter optimization. For the remainder of this work, we select the following hyperparameter values for $\Delta_{\rm OF}$-MLFF: $\{R_{\rm cut}, N_r, L_{\rm max}, \sigma_a \} = \{8, 8, 3, 1\}$, $\{8, 6, 4, 1\}$, and $\{ 8, 6, 4, 1 \}$ for the Al, Si, and Al$_{0.88}$Si$_{0.12}$ systems, respectively. The corresponding values for KS-MLFF are $\{10, 8, 6, 1\}$, $\{10, 8, 5, 1\}$, and $\{10, 8, 5, 1\}$, respectively. The values have been chosen such that larger values of $R_{\rm cut}$, $N_r$, and $L_{\rm max}$, or different values of $\sigma_a$,  do not provide any noticeable increase in the accuracy.

\begin{figure*}[!htbp]
        \includegraphics[width=0.8\linewidth]{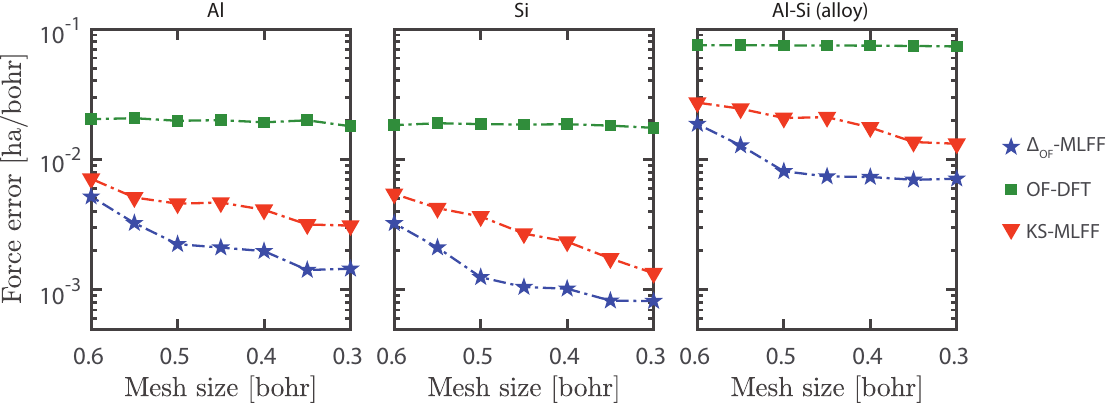}
        \caption{Variation in the force error with mesh size for the 
         Al, Si, and Al$_{0.88}$Si$_{0.12}$ systems, averaged over 1000 MD steps.}
        \label{fig:mesh_error}
\end{figure*}

Next, we study the variation in the $\Delta_{\rm OF}$-MLFF  force errors   with mesh size, the results so obtained are presented and compared with OF-DFT and  KS-MLFF in Fig.~\ref{fig:mesh_error}.  The Kohn-Sham calculations again employ $\Gamma$-point for Brillouin zone integration.  We observe that the $\Delta$-MLFF  errors are not only significantly smaller than KS-MLFF errors for any given mesh size, but are also less sensitive, suggesting that a coarser grid can be used for $\Delta_{\rm OF}$-MLFF relative to KS-MLFF.  In addition, the OF-DFT errors are insensitive to the mesh size, suggesting that the error is completely dominated by the physical error in OF-DFT. The physical error in $\Delta_{\rm OF}$-MLFF appears to be smaller than KS-MLFF, both being significantly lower than OF-DFT, as confirmed below.

\begin{figure*}[!htbp]
        \includegraphics[width=0.8\linewidth]{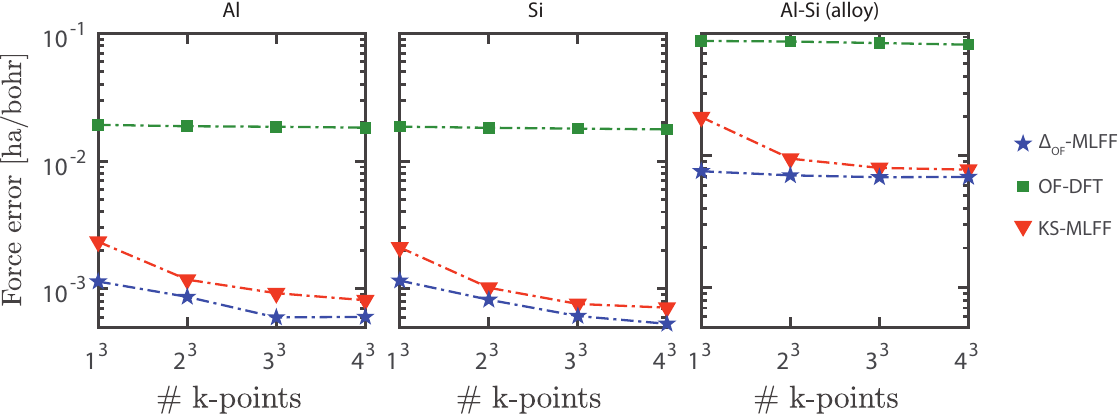}
        \caption{Variation in the force error with number of k-points for the Al, Si, and Al$_{0.88}$Si$_{0.12}$ systems, averaged over 1000 MD steps.}
        \label{fig:kpt_error}
\end{figure*}

Finally, we study the variation in the $\Delta_{\rm OF}$-MLFF  force errors   with number of Monkhorst-Pack wavevectors at which Brillouin zone integration is performed in KS-DFT, i.e., k-points.  The results so obtained are presented and compared with OF-DFT and  KS-MLFF in Fig.~\ref{fig:kpt_error}. As before, both orbital-free and Kohn-Sham calculations employ mesh sizes of $0.35$, $0.35$, and $0.32$ bohr for the Al, Si, and Al$_{0.88}$Si$_{0.12}$ systems, respectively, which translates to the energy and forces being converged to within $10^{-4}$ ha/bohr and $10^{-5}$ ha/atom, respectively. We observe that the $\Delta$-MLFF  errors are not only significantly smaller than KS-MLFF errors for any given number of k-points, but are also less sensitive, suggesting that a coarser k-point grid can be used in $\Delta_{\rm OF}$-MLFF relative to KS-MLFF. In addition, the OF-DFT errors are insensitive to the number of k-points used in KS-DFT, confirming that the error is completely dominated by the physical error in OF-DFT. The physical error in $\Delta_{\rm OF}$-MLFF is confirmed to be smaller than KS-MLFF, both being significantly lower than OF-DFT.

\subsection{\label{Subsec:Accuracy} Accuracy}
We now study the accuracy of the $\Delta_{\rm OF}$-MLFF scheme. To do so, we choose orbital-free and Kohn-Sham parameters, including the mesh size and number of k-points, to be such that the energy and atomic forces are converged to within $10^{-5}$ ha/atom and $10^{-4}$ ha/bohr, respectively. This translates to mesh sizes of 0.35, 0.35, and 0.32 bohr for the Al, Si, and Al$_{0.88}$Si$_{0.12}$ systems, respectively. The corresponding k-point grids are $5 \times 5 \times 5$, $2 \times 2 \times 2$, and $3 \times 3 \times 3$, respectively. We present the results so obtained in Table~\ref{Tab:accuracy} and compare it with those obtained by OF-DFT as well as KS-MLFF. We observe that $\Delta_{\rm OF}$-MLFF reduces the OF-DFT  energy and  force errors by more than two orders of magnitude, bringing it close to the accuracy typically desired in KS-DFT MD simulations. In so doing, it is also more accurate than KS-MLFF, by  $\sim$ 50\%, 50\%, and 20\%   in the energy, and by 35\%, 35\%, and 30\%  in the forces, for the Al, Si, and Al$_{0.88}$Si$_{0.12}$ systems, respectively. Note that the accuracy of both $\Delta_{\rm OF}$-MLFF and KS-MLFF is significantly lower for liquid Al$_{0.88}$Si$_{0.12}$ compared to solid Al and Si. This can be attributed  to the larger configurational space and higher forces in the liquid system. Indeed, the relative accuracy in energy and atomic forces is comparable for all three systems. 

\begin{table*}[!htbp]
\centering
\begin{tabular}{|c|c|c|c|c|c|c|} \hline

      \multirow{2}{*}{Material system} &  \multicolumn{3}{c|}{Energy error [ha/atom] ($\times 10^{-5}$)} & \multicolumn{3}{c|}{Force error [ha/bohr] ($\times 10^{-4}$)} \\
  \cline{2-7}
          &  $\Delta_{\rm OF}$-MLFF &OF-DFT &KS-MLFF & $\Delta_{\rm OF}$-MLFF & OF-DFT  &KS-MLFF  \\
  \hline

    Al & $3.1$ & $776.2$ & $6.2$ & $5.1$ & $170.2$ &  $7.9$ \\
   Si  & $2.8$ & $726.7$ & $5.9$ & $4.2$ &$192.1$  & $6.3$ \\
   Al$_{0.88}$Si$_{0.12}$ & $6.9$ & $2651.3$ & $8.5$&  $61.3$ & $735.9$ & $88.2$ \\
   \hline
\end{tabular}
\caption{Energy and force errors in $\Delta_{\rm OF}$-MLFF, OF-DFT, and KS-MLFF, averaged over 1000 MD steps.}
\label{Tab:accuracy}
\end{table*}

The orders of magnitude improvement of $\Delta_{\rm OF}$-MLFF over OF-DFT is not surprising, given the rather primitive local/semilocal nature of the TFW kinetic energy functional. In particular, even though the electron density from TFW OF-DFT is found to be \emph{close} to KS-DFT,  the energies and consequently the forces are found to be substantially different \cite{thapa2023assessing}. The use of more sophisticated kinetic energy functionals \cite{huang2010nonlocal, constantin2018nonlocal, constantin2019performance, ho2008analytic, karasiev2012issues} is likely to increase the accuracy of not only OF-DFT, but also $\Delta_{\rm OF}$-MLFF.  The ability of the $\Delta_{\rm OF}$-MLFF scheme to achieve higher accuracy than  KS-MLFF can be attributed in part to the smoother and more localized nature of $E_{\Delta}$ relative to $E_{\rm KS}$, which makes  it more amenable to machine learning. Indeed, the use of more advanced kinetic energy functionals in the orbital-free calculations is likely to widen the gap. Notably, the cancellation of long range electrostatic effects in $E_{\Delta}$ is likely to make  $\Delta_{\rm OF}$-MLFF significantly more accurate than KS-MLFF for polar systems. Such systems have not been considered here due to the paucity of accurate and transferable pseudopotentials in OF-DFT. 

\subsection{\label{Subsec:Performance} Performance}
We now study the performance of the $\Delta_{\rm OF}$-MLFF scheme. To do so, we choose all parameters, including mesh size, k-points, and hyperparameters, to be such that the atomic forces  in the $\Delta_{\rm OF}$-MLFF scheme are accurate to within $2 \times 10^{-3}$, $2 \times 10^{-3}$, and $10^{-2}$ ha/bohr for the  Al, Si, and Al$_{0.88}$Si$_{0.12}$ systems, respectively.  To facilitate comparison, the mesh sizes and number of k-points  in KS-MLFF are also chosen to be such that they provide the aforementioned accuracy in the forces, i.e., same accuracy as $\Delta_{\rm OF}$-MLFF.  Note that we do not compare against OF-DFT here since it is not able to achieve this desired accuracy. In fact, the errors in OF-DFT are more than an order of magnitude larger than the accuracy targeted here, as shown in the previous subsection.   
\begin{figure*}[!htbp]       
 \includegraphics[width=0.85\linewidth]{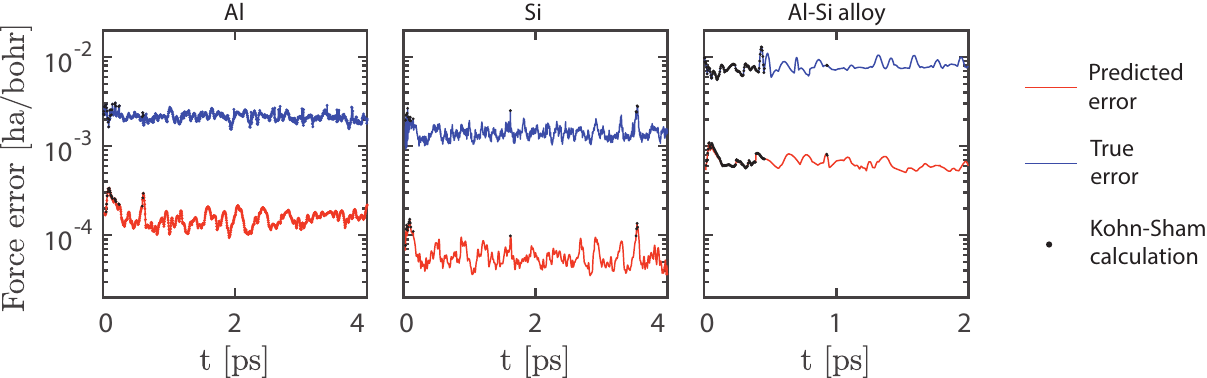}
        \caption{Variation of the true and predicted force error in $\Delta_{\rm OF}$-MLFF for a MD simulation of 1000 steps.  The steps where Kohn-Sham calculations have been performed are also indicated. }
        \label{fig:True_vs_Pred}
\end{figure*}

We present the results so obtained in Table~\ref{Tab:performance}, which also contains the  mesh sizes and number of k-points chosen for each scheme. We observe that both $\Delta_{\rm OF}$-MLFF and KS-MLFF perform Kohn-Sham calculations on only $\sim$1\% of the steps in the  MD simulation containing a total of 10,000 steps,  with the number of such calculations in $\Delta_{\rm OF}$-MLFF being lower than KS-MLFF by $\sim$ 25\%,  25\%, and 15\% for the Al, Si, and Al$_{0.88}$Si$_{0.12}$ systems, respectively. Indeed,  most of the the Kohn-Sham calculations are performed in the beginning of the MD simulation,  with lowering frequency as the simulation progresses, as can be seen in Fig.~\ref{fig:True_vs_Pred}. We also observe that even though the $\Delta_{\rm OF}$-MLFF scheme requires an orbital-free calculation to be performed in each MD step, the overall time is still significantly lower than KS-MLFF, being $\sim$1.5x, $\sim$1.05x, and $\sim$ 3.15x faster for the Al, Si, and Al$_{0.88}$Si$_{0.12}$ systems, respectively. This is because the computational expense is primarily determined by the cost of the Kohn-Sham calculations,  for which the ability to use coarser real space and k-point grids in $\Delta_{\rm OF}$-MLFF, as well as the need to perform fewer such calculations relative to KS-MLFF, can provide substantial savings.  This becomes especially true as the system size increases, due to the cubic scaling nature of Kohn-Sham calculations relative to the linear scaling nature of orbital-free and MLFF calculations.  Note that there is significant scope of improvement in the current OF-DFT implementation,  including the use of preconditioners, which is likely to further increase the efficiency of the $\Delta_{\rm OF}$-MLFF scheme. Also note that though the predicted error and the true error can differ by around an order of magnitude (Fig.~\ref{fig:True_vs_Pred}), they are well-correlated, the correlation coefficient being $0.65-0.70$, which is similar to that obtained for the KS-MLFF scheme, both here and in literature \cite{jinnouchi2019fly}. Indeed, this difference between the true and predicted error is system dependent and is therefore not known apriori. 

\begin{table*}
    \centering
    \begin{tabular}{|c|c|c|c|c|c|c|c|c|} \hline
    
    \multirow{3}{*}{Material system} & \multirow{3}{*}{Method} &\multicolumn{2}{c|}{Discretization parameters}&  \multirow{3}{*}{\parbox[t]{1.8cm}{ \# Kohn-Sham steps\\performed}} & \multicolumn{4}{|c|}{Time [CPU hours]} \\
    \cline{3-4}
     \cline{6 -9}
              &        & Mesh size  & k-point & & \multirow{2}{*}{Kohn-Sham} & \multirow{2}{*}{MLFF} & \multirow{2}{*}{Orbital-free} & \multirow{2}{*}{Total} \\
              &         & [bohr]  & grid & & & & & \\
    \hline
    \multirow{2}{*}{Al}  & $\Delta_{\rm OF}$-MLFF & 0.55 & $1 \times 1 \times 1$ & 66 & 2.9 & 18.1 & 34.2 & 55.2 \\
                         & KS-MLFF & 0.35 & $2 \times 2 \times 2$ & 87 & 63.2 &  18.4 & - & 81.6 \\
    \hline
    \multirow{2}{*}{Si} & $\Delta_{\rm OF}$-MLFF & 0.55 & $1 \times 1 \times 1$ & 55 & 5.5 & 32.1 & 62.2 & 99.8 \\  
                        & KS-MLFF & 0.35 & $1 \times 1 \times 1$ & 76 & 61.4 & 42.7 & - & 104.1 \\
    \hline
    \multirow{2}{*}{Al$_{0.88}$Si$_{0.12}$} & $\Delta_{\rm OF}$-MLFF & 0.55 & $1 \times 1 \times 1$ & 89 & 45.3 & 152.1 & 196.1 & 393.5 \\
                         & KS-MLFF & 0.35 & $2 \times 2 \times 2$  & 103 & 1079.2 & 156.1 & - & 1235.3 \\
    \hline
    
    \end{tabular}
    \caption{Computational time of  $\Delta_{\rm OF}$-MLFF and KS-MLFF for a MD simulation of 10,000 steps.}
        \label{Tab:performance}
    \end{table*}
\subsection{\label{Aggregation}Application: structure of molten Al$_{0.88}$Si$_{0.12}$}

We now apply the $\Delta_{\rm OF}$-MLFF scheme to study the possibility of  Si aggregation in the molten Al$_{0.88}$Si$_{0.12}$ alloy. Experiments have measured a spatial variation in the density of molten Al$_{0.88}$Si$_{0.12}$, which has been attributed to Si aggregation and the formation of Si clusters within the alloy \cite{dahlborg2007structure}. This was however contradicted by KS-DFT simulations \cite{khoo2011ab}, though the relatively small length (500-atom cell) and time scales (10 ps) associated with these simulations has been suggested as a possible reason for the disagreement, which provides the motivation for the current study.

\begin{figure*}[!htbp]
        \includegraphics[width=0.85\linewidth]{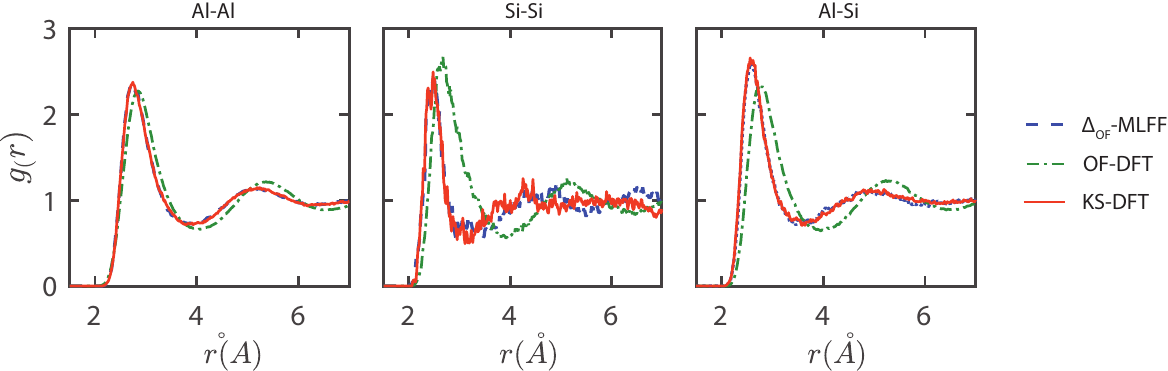}
        \caption{Pair distribution functions for the molten Al$_{0.88}$Si$_{0.12}$ alloy at 1473 K, as computed by $\Delta_{\rm OF}$-MLFF, OF-DFT, and KS-DFT.}
        \label{fig:rdf_compare}
\end{figure*}

First, we verify the scheme's accuracy for this application by calculating the pair distribution functions (PDFs) for the alloy, and comparing it with those obtained by OF-DFT and KS-DFT. As before, we consider a 500-atom cell at a temperature of 1473 K, $\Gamma-$point Brillouin zone integration in Kohn-Sham calculations, 12-th order finite-differences with mesh size of 0.55 bohr, which provides an accuracy of $10^{-4}$ ha/atom and $10^{-2}$ ha/bohr in the energy and atomic forces, respectively, in both orbital-free and Kohn-Sham calculations, and  NVK MD simulations with Gaussian thermostat and a timestep of 2 fs, for a total of 10,000 steps. Note that PDFs are well-converged with respect to all the parameters in the orbital-free and Kohn-Sham calculations,  the results so obtained presented in Fig.~\ref{fig:rdf_compare}. We observe that the PDFs from $\Delta_{\rm OF}$-MLFF  are  nearly indistinguishable to those from KS-DFT, while those from OF-DFT are significantly different, which demonstrates the substantially increased accuracy of the $\Delta_{\rm OF}$-MLFF scheme relative to OF-DFT. Note that the PDFs obtained here  are also in good agreement with those in literature obtained using KS-DFT \cite{khoo2011ab}, further verifying the accuracy of the simulations. 

\begin{figure}[!htbp]
        \includegraphics[width=0.95\linewidth]{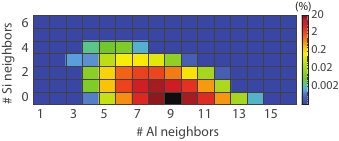}
        \caption{Distribution of  Al and Si neighbors for Si atoms in molten Al$_{0.88}$Si$_{0.12}$  at 1473 K, as computed using $\Delta_{\rm OF}$-MLFF.}
        \label{fig:Neighbour_Dist}
\end{figure}

Next, we perform the MD simulation with a 5000-atom cell for 50,000 steps, i.e., a total time of 100 ps. All  parameters are identical to those used for the calculation of the PDFs. The $\Delta$-MLFF model trained during the 500-atom MD simulation is used as the initial model for the 5000-atom simulation, a strategy referred to as transfer learning. In so doing, the complete simulation requires only six Kohn-Sham calculations, thereby providing significant computational savings. To examine the aggregation of Si atoms, we analyze the distribution of the Al and Si neighbors, as depicted in Fig.~\ref{fig:Neighbour_Dist}. We observe that the ratio of Al to Si nearest neighbors aligns closely with that expected from a random distribution, indicating a homogeneous chemical mixing in the system. In addition, the most common coordination for the Si atoms is nine Al atoms and zero Si atoms in the nearest neighbor shell. Notably, there are very few Si atoms with more than two Si nearest neighbors and virtually none with more than four.  Though these results correspond to 1473 K, we have verified that the qualitative features remain unchanged at  1000 and 1200 K. Indeed, the results are nearly identical to those for the 500-atom cell, which are also in very good agreement with the aforementioned KS-DFT investigation \cite{khoo2011ab}. Therefore, the current results do not support the hypothesis of silicon aggregation \cite{dahlborg2007structure}, in agreement with previous KS-DFT \cite{khoo2011ab} and force field simulations \cite{huang2019liquid}. 

\section{\label{sec:Conclusions}Concluding remarks}
In this work, we have presented a $\Delta$-machine learning model for obtaining Kohn-Sham accuracy from orbital-free DFT calculations. In particular, a MLFF scheme based on the kernel method, with SOAP descriptors and weights determined though Bayesian linear regression, has been employed for capturing the differences  between the Kohn-Sham and orbital-free DFT energies/forces, the resulting formalism referred to as $\Delta_{\rm OF}$-MLFF.  We have implemented $\Delta_{\rm OF}$-MLFF in the context of on-the-fly MD simulations, whose accuracy, performance, and sensitivity to parameters has been studied for the aluminum, silicon, and molten aluminum silicon alloy (Al$_{0.88}$Si$_{0.12}$) systems. We have found that $\Delta_{\rm OF}$-MLFF is not only more accurate than TFW orbital-free DFT by more than two orders of magnitude, but is also more accurate than MLFFs generated solely from Kohn-Sham DFT, while being more efficient and less sensitive to the model parameters. We have applied the $\Delta_{\rm OF}$-MLFF scheme to study the structure of molten Al$_{0.88}$Si$_{0.12}$, for which we have found no evidence of Si aggregation, in agreement with a previous Kohn-Sham DFT investigation  performed at an order of magnitude smaller length and time scales. 

The use of more advanced kinetic energy functionals for the orbital-free calculations in the $\Delta_{\rm OF}$-MLFF scheme is likely to further increase its accuracy. In this context, the choice/development of functionals that have the best balance between accuracy and computational cost is a worthy subject for future research. The implementation of $\Delta_{\rm OF}$-MLFF on GPUs is likely to significantly bring down the solution times, as demonstrated recently for the associated Kohn-Sham calculations \cite{sharma2023gpu}, making it another subject worthy for future research. From a MLFF perspective, the current findings suggest that orbital-free DFT and other fast physical approximations can provide a valuable complement to machine learning techniques, indicating that renewed focus on improving the speed, accuracy, and general applicability of orbital-free DFT is warranted.


\begin{acknowledgments}
The authors gratefully acknowledge the support of the U.S. Department of Energy, Office of Science under grant DE-SC0023445. This work was performed, in part, under the auspices of the U.S. Department of Energy by Lawrence Livermore National Laboratory under Contract DE-AC52-07NA27344. This research was also supported by the supercomputing infrastructure provided by Partnership for an Advanced Computing Environment (PACE) through its Hive (U.S. National Science Foundation through grant MRI-1828187) and Phoenix clusters at Georgia Institute of Technology, Atlanta, Georgia. P.S. acknowledges discussions with Igor I. Mazin regarding the theoretical underpinnings of orbital-free and Kohn-Sham DFT.  
\end{acknowledgments}

\appendix
\section{\label{App:StressTensor} Stress tensor in the $\Delta$-machine learning model}
In the $\Delta$-machine learning model, the correction to the stress tensor computed by orbital-free DFT  takes the form:
\begin{align}
\sigma_{\Delta, \lambda \mu}& = \frac{1}{V}\left.\frac{\partial E_{\Delta}^{F}}{\partial F_{\lambda \mu}}\right\vert_{\mathbf{F}=\mathbf{I}} \, , \nonumber \\
& = -\frac{1}{V}\sum_{e=1}^{N_{e}} \sum_{i=1}^{N_A^e} \sum_{t=1}^{N_{T}^{e}} \tilde{w}_{t}^{e} \xi k\left( \mathbf{x}_{i}^{e}, \tilde{\mathbf{x}}_t^{e} \right) \mathbf{z}_{i,t}^e \cdot  \left.\frac{\partial {\mathbf{x}_{i}^{e}}^{F}}{\partial  F_{\lambda \mu} }\right\vert_{\mathbf{F}=\mathbf{I}} \,.
\label{Eq:StressFinalML}
\end{align} 
where 
\begin{align} 
\left. \frac{\partial[{\mathbf{x}_i^e}^{F}]_{n_1 n_2 l}^{e_1 e_{2}}}{\partial   F_{\lambda\mu} }\right\vert_{\mathbf{F}=\mathbf{I}} &= c_l \sum_{m=-l}^{l} \left( \frac{\partial {c^{e}}^{F}_{i, e_1,  n_1 lm}}{\partial F_{\lambda\mu}} {c^{e *}}^{F}_{i, e_{2}, n_2 lm} \right. \nonumber\\
&+ \left.\left.{c^e}^{F}_{i, e_1, n_1 lm} \frac{\partial  {c^{e *}}^{F}_{i, e_{2}, n_2 lm}}{\partial   F_{\lambda\mu}}  \right)\right\vert_{\mathbf{F}=\mathbf{I}}  \,,
\label{Eq:DescriptorDerivF}
\end{align}
with 
\begin{small}
\begin{align} 
\left.\frac{\partial {c^{e}}^{F}_{i, \Tilde{e}, nlm}}{\partial F_{\lambda \mu} }\right\vert_{\mathbf{F}=\mathbf{I}} = & \sum_{j=1}^{N_A^{\Tilde{e}}}  \frac{(r_{j\lambda}^{\Tilde{e}}-r_{i\lambda}^{e})(r_{j\mu}^{\Tilde{e}}-r_{i\mu}^{e})}{|\mathbf{r}_{j}^{\tilde{e}} - \mathbf{r}_i^e|} h_{nl}' \left( |\mathbf{r}_{j}^{\tilde{e}} - \mathbf{r}_i^e|\right) Y_{lm}^* \left( \mathbf{r}_{j}^{\tilde{e}} - \mathbf{r}_i^e \right)  \nonumber \\
& + h_{nl} \left(|\mathbf{r}_{j}^{\tilde{e}} - \mathbf{r}_i^e| \right)\frac{\partial Y_{lm}^* \left( \mathbf{r}_{j}^{\tilde{e}} - \mathbf{r}_i^e \right)} {\partial (\mathbf{r}_{j}^{\tilde{e}} - \mathbf{r}_i^e)} \cdot \mathbf{J}^{\lambda \mu} (\mathbf{r}_{j}^{\tilde{e}} - \mathbf{r}_i^e)\,.
\label{Eq:CnlmDerivF}
\end{align}
\end{small}
Above, $V$ is the volume of the simulation cell, $F_{\lambda \mu}$ are the Cartesian components of the deformation gradient tensor $\mathbf{F}$, $(\cdot)^F$ denotes the quantities after deformation,   and  $\mathbf{J}^{\lambda \mu} \in \mathbb{R}^{3\times 3}$ denote the single-entry matrices that have a value of 1 at the $(\lambda,\mu)$ position and zeros elsewhere.

\section*{Data Availability Statement}
The data that support the findings of this study are available within the article and from the corresponding author upon reasonable request.

\section*{References}
\bibliography{Manuscript}

\end{document}